\documentclass[aps,pre,twocolumn,showpacs,showkeys,a4paper]{revtex4-1}
 \usepackage{graphicx}
  \usepackage{amsmath}
 \usepackage{amssymb}
 \usepackage{amscd}
 \usepackage{color}

\begin{document} 

 \title{Exact Current Statistics of the ASEP with Open
   Boundaries}
  \author{Mieke Gorissen$^{(1)}$, Alexandre Lazarescu$^{(2)}$, Kirone
   Mallick$^{(2)}$ and Carlo Vanderzande$^{(1,3)}$}
 \affiliation{(1) Faculty of Sciences, Hasselt University, 3590 Diepenbeek, Belgium\\
 (2) Institut  de Physique
   Th\'eorique, C. E. A.  Saclay, 91191 Gif-sur-Yvette Cedex, France\\
  (3) Instituut Theoretische Fysica, Katholieke Universiteit Leuven, 3001
  Heverlee, Belgium}
  \pacs{05-40.-a; 05-60.-k; 02.50.Ga}

 \begin{abstract} 
Non-equilibrium systems are often characterized by the transport of
some quantity at a macroscopic scale, such as, for instance, a current
of particles through a wire. The Asymmetric Simple Exclusion Process (ASEP)
is a paradigm for non-equilibrium transport that is amenable to  exact
analytical solution. 
 In the present work, we
 determine the full statistics of the
current in the finite size open ASEP for all values of the
parameters. Our exact  analytical results are checked against numerical
calculations using DMRG techniques.
  \end{abstract}

\maketitle 

A  system containing carriers (of thermal energy, mass, or
electrical charge)  and  subject to a
driving field in its bulk, or to a difference of potentials between
its boundaries,  will usually evolve to a non-equilibrium  steady
state (NESS) with  a non-vanishing  macroscopic
current of heat, particles or charges flowing  through it.  
Due to  the presence of this macroscopic current,
time-reversal invariance is violated. This is a  situation  which
lies   beyond  the realm of  traditional thermodynamics:
steady-state transitions at the  microscopic level break detailed
balance and the principles  of  equilibrium statistical mechanics do
not apply.  Hence, for a system that is bulk-driven, boundary-driven,
or both, no suitable generalization of the Gibbs-Boltzmann formalism
exists that would allow us to  predict the value of the current
and of its fluctuations from  first principles.

 During the last two decades, substantial progress has been made
 towards a  statistical theory of  non-equilibrium   systems
 \cite{DerridaCairns,Gallavotti,LebSpohn,Zia,Schutzrev,Spohn}. Large
 deviation functions, that encode atypical fluctuations   of a
 physical observable, are likely to be  the best candidates to generalize
 the traditional thermodynamic potentials.
  Moreover, it has been proved 
  that  large deviations functions display  symmetry properties,
  called  `Fluctuation
 Theorems', that remain  valid far from equilibrium 
 \cite{Gallavotti}. These remarkable
 relations imply linear response theory in the vicinity of
 equilibrium. Hence, the determination of 
 large deviations in a non-equilibrium
 system, whether theoretically, numerically, or  experimentally,
  is  a  question of fundamental importance
  \cite{Bertini,Bodineau,DLeb,DLSpeer,Gorissen,Kurchan,Takeuchi,Touchette}.

  There are very few  models in  non-equilibrium physics that 
  can be studied analytically. Among these,  the
  asymmetric simple exclusion process (ASEP) has become  a
  paradigm \cite{DerridaRep,KLS,Varadhan}.  
  The ASEP is a one-dimensional lattice gas  model in which 
  particles perform biased random walks and interact through an
  exclusion constraint that mimics a hard-core repulsion:  two
  particles cannot occupy the same site at a given time.
  This minimal system appears as a building block in a great variety
  of phenomena that involve low-dimensional transport with
  constraints. Invented  originally  to represent the motion of
  ribosomes along mRNA during protein synthesis,
  this  model  plays  a seminal role in   non-equilibrium
  statistical mechanics and has  been applied to problems as different as
  surface growth, biological transport,
  traffic flow and pure mathematics 
  \cite{Chou,Johansson,KPZ,Krug,SasamSpohn,Schutzrev}.

  In the long time limit, the ASEP reaches a NESS with  a fluctuating
  macroscopic   current.  Exact
  results have been derived  for the exclusion process 
  on a  periodic ring  and
  on the infinite line, using  the Bethe Ansatz,
  determinantal processes and random matrix theory
  \cite{Dhar,Sylvain4,SasamSpohn,SchutzDet,TracyWidom}.
  For  open boundaries, the steady-state
   has a recursive structure  \cite{DeDoMuk}
  that can be encoded by a 
  matrix  product representation \cite{DEHP}, a fruitful method
  to analyze  low-dimensional transport models  \cite{MartinReview,Schutzrev}.
  The mean value  of the stationary  current, the associated  density
  profiles and the phase diagram of the open ASEP are known exactly
  \cite{DeDoMuk,DEHP}.  However, 
  finding the full  statistical properties of the  current in the open ASEP  
   has remained, until now,  an outstanding  challenge that 
  has stimulated many works
  \cite{Gorissen,Bodineau,BoDerr,DEMal,Doucot,deGierNew,DWT,SasaPasep1}. 
   A recent  conjecture based on the Bethe Ansatz  \cite{deGierNew}  
  gives  the asymptotic  behavior of the large
  deviation function of the current  for infinitely large systems in
  some specific  regions of the phase diagram. 
  In the present work, we give exact analytic expressions
  for the full statistics of the current, that are valid for arbitrary
  system sizes and boundary parameters, thus solving this
  long-standing  problem. 

\begin{figure}[ht]
\begin{center}
 \includegraphics[width=0.5\textwidth]{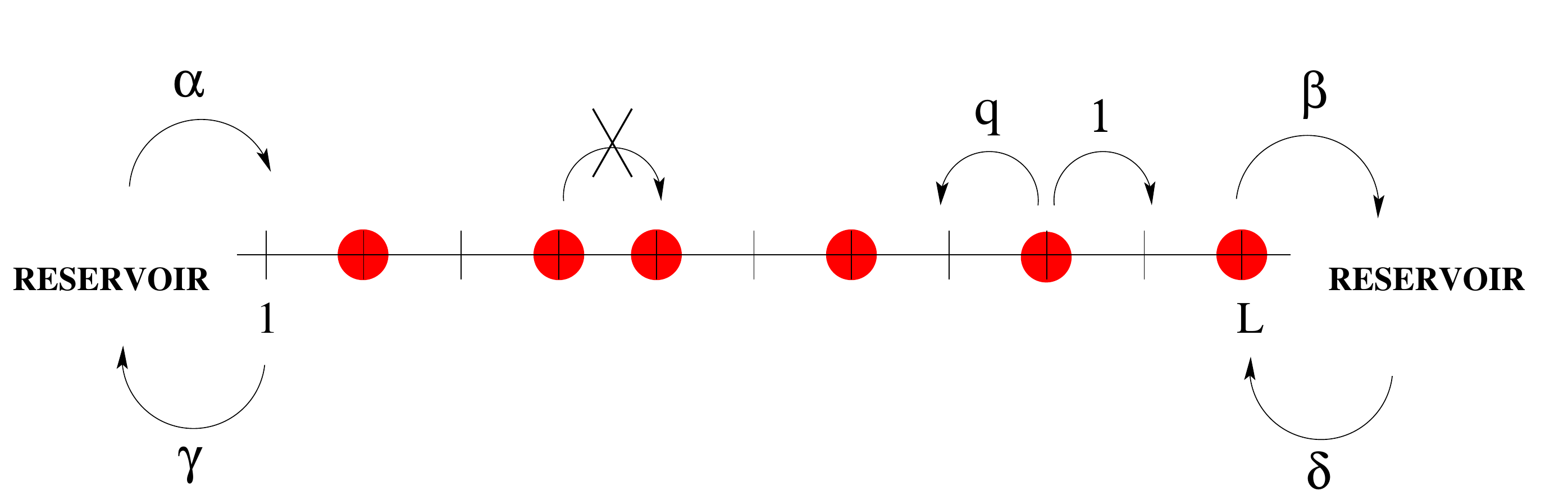}
  \caption{Dynamical rules for the ASEP  with open boundaries. The rate
 of forward jumps has been normalized to 1. Backward jumps occur with rate
 $q < 1$. All other parameters are arbitrary.}
\label{fig-PASEP}
 \end{center}
 \end{figure}

 The dynamics of the  ASEP  is that of a continuous time Markov chain:
  during an infinitesimal time interval $dt$,
  a particle located on a site can jump forward
 to the next adjacent site with  rate 1
 and hop backward to the previous  site with  rate $q$, provided 
 these sites are empty. A particle can enter
 the  site $1$ with rate $\alpha$ and the  site $L$ with rate $\delta$, and can exit
  from the site $1$ with rate $\gamma$ and from the site $L$ with rate $\beta$ 
  (see Fig.~\ref{fig-PASEP}). 
 Each of the $2^L$ microscopic configurations 
 ${\mathcal C}$  of the ASEP can be  written as a binary string of length $L$,
 $(\tau_1, \ldots, \tau_L)$, where  $\tau_i = 1$ if the site $i$ is
 occupied and  $\tau_i = 0$ otherwise.  The probability
 $P_t({\mathcal C})$ of being in configuration ${\mathcal C}$ at time
 $t$ evolves  according to the Master equation:
\begin{equation}
 \frac{d P_t({\mathcal C})} {dt} = \sum_{{\mathcal C'}} M({\mathcal
   C}, {\mathcal C'}) P_t({\mathcal C'}) \, .
 \label{Eq:Markov}
\end{equation}
 The non-diagonal  matrix element $M({\mathcal C}, {\mathcal C'})$
 represents the transition rate from ${\mathcal C'}$ to ${\mathcal
   C}$.  The diagonal element  $M({\mathcal C}, {\mathcal C}) = -
 \sum_{{\mathcal C'} \neq  {\mathcal C}} M({\mathcal C'}, {\mathcal
   C}) $   is equal to  minus  the exit rate from ${\mathcal C}$.

 In the long time limit, the ASEP reaches a NESS where
 each configuration ${\mathcal C}$   occurs with a
 probability $P^\star({\mathcal C})$, that  
  can be written as  a matrix product   \cite{DEHP}:
\begin{equation}
  P^\star(\mathcal C) = \frac{1}{Z_L} \langle W | \prod_{i=1}^L  \left(
  \tau_i { D} + (1 - \tau_i)  { E}\right) | V  \rangle \, ,
\label{MPA}
\end{equation}
 where  the operators $ { D}$ and  $ { E}$, the bra-vector $
 \langle W | $ and the ket-vector $ | V  \rangle$ satisfy  quadratic
 algebraic relations 
\begin{eqnarray}
             { D} \,  { E} -q { E}  { D} & = &  (1 -q )\left(  { D} \,
             + \, { E}  \right)  \nonumber \\ ( \beta { D} - \delta
             { E})  \,  | V  \rangle   & = &  (1 -q )  | V  \rangle
             \nonumber   \\ \langle W | \, ( \alpha { E} - \gamma
                         { D})  & = &   (1 -q ) \langle W |  \,\, . 
 \label{DEHPAlgebra}
\end{eqnarray}
The normalization constant in equation~(\ref{MPA}) is given by
 $ Z_L =  \langle W | \left( { D} + { E} \right)^L | V  \rangle\,.$
  The matrix product representation allows us to   determine  stationary equal-time
 correlations and  density profiles for any  system size $L$.  
 
 For $L \to \infty$,  the ASEP has  three
 phases whose boundaries are given in terms of the effective
 densities $\rho_a=1/(a_++1)$ and $\rho_b=b_+/(b_++1)$ of
  the left and right reservoirs, where
\begin{eqnarray}
a_{\pm}  &=&  \frac{ (1-q-\alpha+\gamma)\pm
  \sqrt{(1-q-\alpha+\gamma)^2+4\alpha\gamma}}{2\alpha} \, , \\ b_{\pm}
&=&  \frac{(1-q-\beta+\delta) \pm
  \sqrt{(1-q-\beta+\delta)^2+4\beta\delta}}{2\beta}\, .
\end{eqnarray}
The ASEP
is in the Maximal Current (MC) phase when $\rho_a> 1/2$
 and  $\rho_b <1/2$, in the Low Density (LD) phase when   $\rho_a < 1/2$ and   
 $ \rho_a + \rho_b < 1 $ and
 in the High Density (HD) phase when  $\rho_b >1/2$ and
 $ \rho_a + \rho_b > 1 $ .

 For a system of size $L$,  the average value  ${J}$ of the
 stationary  current   is  given by the ratio $Z_{L-1}/Z_L$ which can
 be expressed in terms of orthogonal polynomials \cite{SasaPasep1}.
 However, the
 fluctuations  of the steady-state  current can not be  calculated
 from the knowledge of the  stationary probabilities alone. 
 In order to study  the current we introduce an  observable 
 $Y_t$ that counts the number of  particles  exchanged between
 the system and  the left  reservoir
 between times  $0$ and $t$. Therefore,   $Y_{t +dt} = Y_t +y$  where $y= +1$
  if a particle enters the  site 1,
 $y= -1$ if  a particle  exits  from 1 during the interval $dt$ 
  and $y= 0$ otherwise.  These three mutually  exclusive types of
 transitions lead to a three parts  decomposition of  the generator $M$:
  $M = M_{+} +  M_{-} + M_{0}$.  We note that  $Y_t$ also 
 represents the total integrated  current that has flown through the system
  till  time  $t$. 
  When $t \to \infty$, the expectation value of 
  $Y_t/t$ converges to
 the average stationary current  ${J}$. The convergence rate is quantified
 by the large deviation function $\Phi(j)$, characterizing 
 non-typical fluctuations of  $Y_t$ and  defined as 
 $P\left(\frac{Y_{t}}{t}=j\right) {\sim}e^{-t \Phi(j)}$.

 A different manner  to encode 
 the statistics  of $Y_t$ is  through  its  characteristic  function
 which, in the long time limit,  behaves as
 $  \langle e^{\mu \, Y_t} \rangle \simeq e^{{\mathcal E}(\mu) t } \, ,$ 
 where  ${\mathcal E}(\mu)$  is the cumulant generating function of 
 $Y_t$, and is  the  Legendre  transform of the large-deviation function
   $\Phi(j)$  \cite{Touchette}:
$  {\mathcal E}(\mu) = \max_j \left( \mu j -  \Phi(j)    \right)$.
 Following \cite{DLeb, LebSpohn}, one can prove that 
 ${\mathcal E}(\mu)$ is the largest eigenvalue
 of the deformed operator 
 $M(\mu) =  \rm{e}^\mu M_{+} +  \rm{e}^{-\mu} M_{-} + M_{0}$. Thus, the 
 calculation of the cumulants of the current 
 is equivalent to  an eigenvalue problem.

  For the ASEP with periodic boundary conditions, 
  $M(\mu)$ can be diagonalized by Bethe Ansatz, leading 
  to a full  solution for the current fluctuations \cite{DLeb,Sylvain4}.
  In the case of open boundary conditions, integrability conditions
  are only met on hypersurfaces of the parameter space and 
  the Bethe Ansatz can  be used  only for $L \to \infty$
  and in specific regions of the phase diagram \cite{deGierNew}.

  We have obtained a solution valid for all parameters and all system sizes
   using a  generalized matrix product representation.
  The components
  $F_{\mu}({\mathcal C})$ of the  dominant eigenvector  $F_{\mu}$ of $M(\mu)$ can
  be  expanded formally   as a power-series with respect to 
 $\mu$  to any given  order $k \ge 0$.  For each value of $k$, we have
  proved  rigorously  \cite{LazAl} that  $F_{\mu}$ can be represented by 
  a matrix product   Ansatz  up to corrections of order $\mu^{k+1}$ i.e., 
 \begin{equation}
F_{\mu}({\mathcal C}) =     \frac{1}{Z_L^{(k)}} \langle W_k | \prod_{i=1}^L  \left(
  \tau_i {D_k} + (1 - \tau_i)  {E_k}\right) | V_k  \rangle 
+ {\mathcal O}\left(\mu^{k+1}\right)
\label{MPA2}      
\end{equation}
The matrices  $D_k$ and $E_k$ are constructed  recursively starting with
 $D_1 =D$ and $E_1 =E$ and 
 \begin{eqnarray}
       D_{k+1} &=& 
 (1 \otimes 1 + d \otimes e ) \otimes  D_k
 + ( 1 \otimes d   +  d \otimes 1)  \otimes  E_k  
   \nonumber  \\
       E_{k+1}  &=&  
   (1 \otimes 1 +  e   \otimes d )  \otimes  E_k 
+ (e  \otimes 1 + 1 \otimes  e )  \otimes   D_k  \nonumber \\
\end{eqnarray} 
where we have defined the operators
  $d  = D-1$ and $e = E-1$ that 
 satisfy the $q$-deformed harmonic oscillator algebra
 $de  - q ed  = 1 - q$. These matrices are related to the ones used for the matrix product solution of the multispecies periodic ASEP \cite{Multi}. 

The boundary vectors $\langle W_k | $   and   $ | V_k  \rangle  $
are  constructed by taking tensor products of bra and ket vectors.
 We start with    $| V_1  \rangle  =   | V \rangle  $
 and $\langle W_1 |      =  \langle W|$ and iterate
 \begin{eqnarray} 
   | V_{k+1}  \rangle  &=&   | V \rangle\otimes  |\tilde{V} \rangle \otimes
  | V_{k}  \rangle  \\
  \langle W_{k+1} |    &=& \langle W^\mu| \otimes\langle  \tilde{W}^\mu |
   \otimes \langle W_{k} |   
    \,,
\end{eqnarray} 
 where $| V \rangle $ is defined in Eq.~(\ref{DEHPAlgebra}) and 
 \begin{eqnarray} 
     [ \beta ( 1  - d)  
   - \delta ( 1  -  e ) ] \,  | \tilde{V}   \rangle  &=& 0    \\
       \langle W^\mu| [ \alpha(1 + {\rm e}^\mu \,  e )
  - \gamma (1 +  {\rm e}^{-\mu} \,  d )]
    &=& ( 1 - q)   \langle W^\mu|    \\
       \langle  \tilde{W}^\mu | [  \alpha(1 - {\rm e}^\mu \,  e )
  - \gamma (1 -  {\rm e}^{-\mu} \, d )] &=& 0 \, .
 \end{eqnarray} 

 This matrix Ansatz allows  us to calculate the cumulants to any desired
 order $k$. 
  Our  central result  is a  parametric formula  for  the cumulant
 generating function  ${\mathcal E}(\mu)$:   
 \begin{eqnarray}
           \mu = - \sum_{k \ge 1}  C_k \frac{B^k}{k} \,   \hbox{ and } 
  {\mathcal E}  =-(1   - q)\sum_{k \ge 1}   D_k \frac{B^k}{k}  \, , 
\end{eqnarray}
where  $B$ is a formal   parameter that has to be eliminated from  the
 two equations. We  emphasize that 
 similar parametric expressions have appeared in all
 known exact expressions for the current cumulant  generating function
  \cite{DLeb,LazarescuMallick,Sylvain4}   and
  a similar  generic form   was derived from
 the additivity principle in \cite{Bodineau}.
 The function  ${\mathcal E}(\mu)$ is fully specified from the knowledge of the 
 scalars  $C_k$  and $D_k$. These  are  given by contours integrals in the
 complex plane along a  contour ${\Gamma}$ (to be defined below): 
 \begin{eqnarray}
          C_k =   \oint_{\Gamma} \frac{dz}{2\, i\, \pi}
          \frac{\phi_k(z)}{z}  \, ,  \quad
           D_k  =  \oint_{\Gamma} \frac{dz}{2\, i\, \pi} 
           \frac{\phi_k(z)}{(z+1)^2} \, . \label{valeur:CkDk} 
\end{eqnarray}
  The  $\phi_k(z)$'s are obtained as follows:  we define a function $W_B(z)$
 that depends on the   parameter $B$ 
 \begin{eqnarray}
 W_B(z) =  \sum_{k \ge 1}  \phi_k(z)  \frac{B^k}{k} \, ,
\end{eqnarray}
 and we find  that  $W_B(z)$ 
  is uniquely determined as the solution of the 
 functional equation:
\begin{equation}
  W_B(z)=-\frac{1}{2} \ln\Bigl(1-B F(z) e^{X[W_B](z)}\Bigr) \,\, ,
\label{eq:W}
\end{equation}
  where  $F(z)$ is given by the expression
\begin{equation}
\label{FPASEP}
\frac{(1+z)^L(1+z^{-1})^L(z^2)_{\infty}(z^{-2})_{\infty}}
{
 (a_{+}z)_{\infty}     (\frac{a_{+}}{z})_{\infty}   
(a_{-}z)_{\infty}       (\frac{a_{-}}{z})_{\infty}   
(b_{+}z)_{\infty}      (\frac{b_{+}}{z})_{\infty}   
(b_{-}z)_{\infty}     (\frac{b_{-}}{z})_{\infty}   
 }
\end{equation}
 with
 $ (x)_{\infty}=\prod_{k=0}^{\infty}(1-q^k x) \, .$ 
 We note that   $F(z)$ appears  in the definition of the Askey-Wilson
polynomials, known to be relevant to the open ASEP \cite{SasaPasep1}.
 The operator $X$ is  a  linear  integral operator:
\begin{equation}\label{X}
X[W_B](z_1)=\oint_{\Gamma} \frac{d{z_2}}{2 \imath \pi \,
  {z_2}}W_B({z_2})K\left(\frac{z_1}{z_2} \right) \, , 
\end{equation}
where the kernel  $K$ is   given by
\begin{equation}\label{Kbis}
K(z)=2\sum_{k=1}^{\infty}\frac{q^k}{1-q^k}
  \, \left\{ z^k
 + z^{-k}\right\}
\end{equation}
 and the contour  ${\Gamma}$ in the complex plane   encircles (once)  the points 0,
 $q^k a_{+},q^k a_{-},q^k b_{+}$ and $q^k b_{-}$   for all integers  $k \ge 0$.
 The kernel $K(z_1/{z_2})$ was used in \cite{Sylvain4} to calculate the current
 fluctuations in the  periodic case.

 The functional equation~(\ref{eq:W}) contains the complete
  information about the  current statistics: by 
 solving it  iteratively to any order $k$,  we obtain
  the   first  $k$ cumulants of the current. At first order, we have
 $\phi_1(z) = F(z)/2$ and  the mean value  of the current is
\begin{equation}
 J =  \lim_{t \to \infty} \frac{\langle  Y_t\rangle }{t} =  (1 -q) \frac{D_1}{C_1}
=  (1 -q) \frac{ \oint_{\Gamma} \frac{dz}{2\, i\, \pi}
          \frac{F(z)}{z}}{\oint_{\Gamma} \frac{dz}{2\, i\, \pi}
          \frac{F(z)}{(z+1)^2}}
\end{equation}
This expression is identical to that  given  in \cite{SasaPasep1}.
  At second order,
the variance of the current is:
\begin{equation}
 \Delta =  \lim_{t \to \infty} \frac{\langle Y_t^2\rangle - \langle  Y_t\rangle^2  }{t} = 
 (1 -q)  \frac{D_1 C_2 - D_2 C_1}{2 C_1^3}
\label{eq:DiffConst}
\end{equation}
where $C_2$ and $D_2$ are obtained using (\ref{valeur:CkDk}) with
\begin{equation}
\phi_2(z)= \frac{1}{2}\biggl( F^2(z)+
\oint_{\Gamma}
 \frac{dz_{2} F(z)F(z_{2})K(z/z_{2})}{2\imath\pi z_{2} } \biggr).
  \nonumber
\end{equation}
For higher cumulants,  exact expressions  similar to Eq.~(\ref{eq:DiffConst})
 are obtained  and can be expressed via a
  combinatorial tree expansion akin to that found in
 the periodic case \cite{Sylvain4}.  The expression of the diffusion constant 
 $\Delta$ generalizes the  formula  of \cite{DEMal} 
 obtained for the  totally asymmetric exclusion process (TASEP)
 in which   $q = \delta = \gamma =0$.  For the TASEP, 
the kernel  $K$ and the operator X   vanish  identically and  $F(z)$ reduces to
\begin{equation}
  F_{{\rm TASEP}}(z) =\frac{ - ( 1 + z)^{2L} ( 1 - z^2)^2 }{z^L (1 -az)(z -a)(1 -bz)(z -b)}
\end{equation}
 with   $a = \frac{ 1 - \alpha}{ \alpha}$  and  $b =  \frac{ 1 - \beta}{ \beta}$;
 then,  Eq.~(\ref{eq:W}) leads to 
 $\phi_k(z) = F_{{\rm TASEP}}^k(z)/2$ 
 and  the  results of \cite{LazarescuMallick} are  retrieved.
 For a  periodic system  of size $L$ with $N$ particles,
 the current fluctuations   can be brought
 into  the framework described here with the same generalized  matrix Ansatz,
 but   the  boundary vectors are replaced  by a trace and  
 equation~(\ref{eq:W})  is modified as follows:
 the prefactor 1/2 is removed,
 $F(z) = (1 +z)^L/z^N$ and the Kernel $K$ is still given by (\ref{Kbis}).
 Then, the results of  \cite{Sylvain4}, originally  obtained  by Bethe Ansatz, 
 are retrieved.

\begin{figure}[ht]
\begin{center}
 \includegraphics[height=5.5cm]{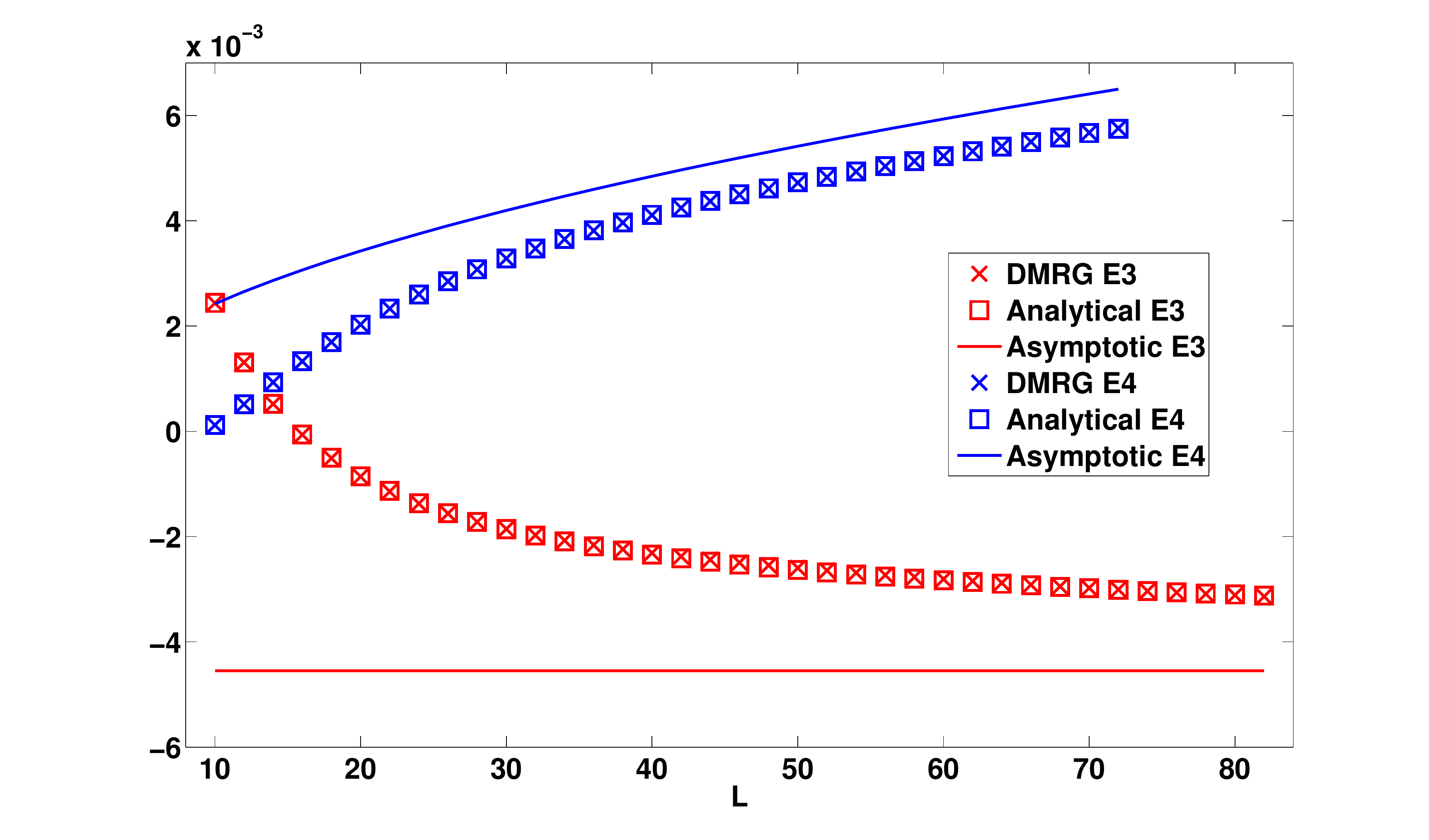} 
  \caption{Third (E3, red) and fourth (E4, blue)
 cumulants in the Maximal Current phase,
 with $q=0.5$, $a_+=b_+=0.65$, $a_-=b_-=0.6$; the  full 
 lines represent the corresponding
 large size asymptotic  behaviors.}
\label{E3E4MC}
 \end{center}
 \end{figure}

\begin{figure}[ht]
\begin{center}
 \includegraphics[height=5.5cm]{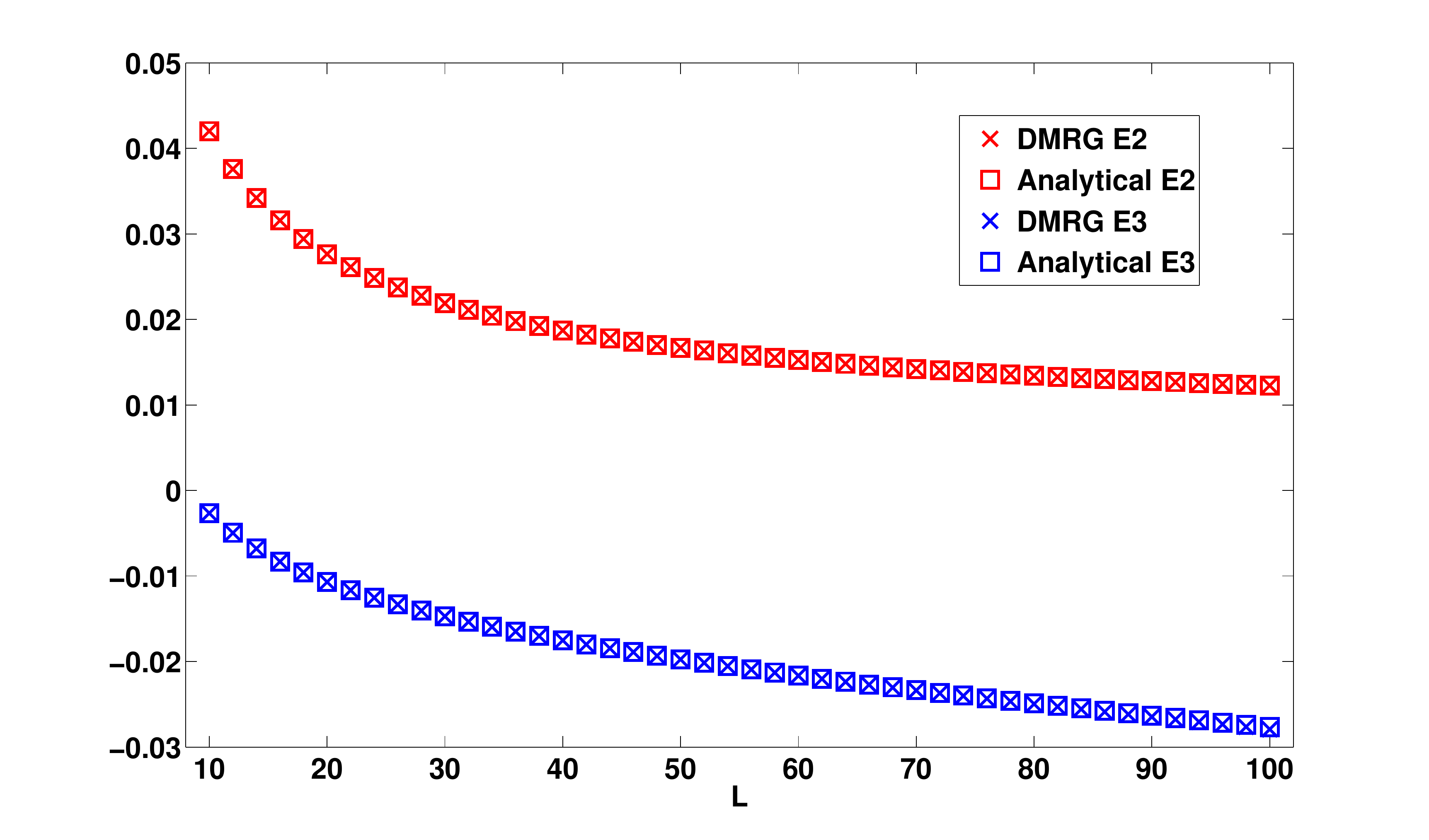} 
  \caption{Second (E2, red) and third (E3, blue) cumulants in the High Density phase, with $q=0.5$, $a_+=0.28$, $b_+=1.15$, $a_-=-0.48$ and $b_-=-0.27$.}
\label{E2E3HD}
 \end{center}
 \end{figure}

The  derivation of the  above results involves combinatorial
identities for   matrix elements of  the  generalized  matrix Ansatz.
Some of these identities were guessed by induction rather
 than  mathematically proved \cite{LazAl}. 
It was therefore necessary to validate our calculations numerically.
  For  small size systems ($L \le 7$),  expressions for the cumulants  
  have been checked against the exact values from direct  calculations. For larger
 systems ($L \le 100$), we compared the analytical formulas with  numerical computations 
 of the cumulants performed using a DMRG
 method. That method, originally introduced to study ground state properties of quantum spin chains \cite{White} was recently adapted to calculate the highest eigenvalues of deformed stochastic operators like $M(\mu)$   \cite{Gorissen}. A few of those results are displayed in
 Figs~\ref{E3E4MC} and \ref{E2E3HD}.

  For  large system sizes, $L \to \infty$, 
  the cumulant generating function takes different expressions in 
different phases. These are derived from
  an asymptotic analysis involving the leading singularities of $F(z)$
\cite{MartinReview,SasaPasep1,LazAl}.
  In the  Low Density  phase, the dominant singularity is the  pole at $a_{+}$
  leading to  $ \phi_k(z) \sim F^k(z)$. Using the Lagrange inversion formula
  as in \cite{LazarescuMallick}, we obtain
   \begin{equation}
  {\mathcal E}(\mu) =  (1 -q) (1-\rho_a)
   \frac{ {\rm e}^\mu -1 }{{\rm e}^\mu +(1-\rho_a)/\rho_a }.
\end{equation}
This expression agrees with the one found in
 \cite{deGierNew} using the Bethe Ansatz. Its Legendre Transform matches
 the prediction   of the additivity principle
 \cite{BoDerr}:
 \begin{equation}
 \Phi(j) =(1 -q) \left\{ \rho_a - r  + 
  r ( 1 - r) \ln \left( \frac{1 - \rho_a}{\rho_a}
  \frac{r}{1-r} \right) \right\} \, ,  
\end{equation}
 where the current $j$ is parametrized as  $j = (1 -q) r ( 1 - r) .$
 The HD Phase leads to similar expressions with $ a_{+} \to b_{+}$
 and  $\rho_a \to  1 - \rho_b$. In both cases, the statistics of the current do not depend on system size  in the large $L$ limit.

In the  Maximal Current phase, we find that
  $k$-th cumulant grows as $L^{(k-3)/2}$. When  $L \to \infty$, we have 
 \begin{eqnarray}
           \mu &=&-\frac{L^{-1/2}}{2\sqrt{\pi}}\sum_{k=1}^{\infty}\frac{(2k)!}{k!k^{(k+3/2)}}B^k \, ,
   \\ {\mathcal E}-\frac{1-q}{4}\mu  &=&-\frac{(1-q)L^{-3/2}}{16\sqrt{\pi}}\sum_{k=1}^{\infty}\frac{(2k)!}{k!k^{(k+5/2)}}B^k  \, .
\end{eqnarray}
 These expressions  have the same structure as those obtained  
 for the case  of  a periodic ring \cite{DLeb} and  the large deviation
 functions have the same asymptotic behavior.
 Moreover, for  the  open TASEP of size $L$  with $\alpha=1$ and $\beta=1/2$,
 we observed that the formulas are {\it identical} to  those for 
  the half-filled   periodic TASEP
 of size $2L+2$. 

 Along the shock line ($\rho_a=1-\rho_b<1/2$), we obtain 
 \begin{eqnarray}
           \mu &=&-2L^{-1}\frac{(1+a_+)}{(1-a_+)}\sum_{k=1}^{\infty}\frac{k^{2k-1}}{(2k)!}B^k \, ,
   \\ {\mathcal E}-\frac{(1-q)a_+}{(1+a_+)^2}&\mu&  =-2L^{-2}\frac{(1-q)a_+}{(1-a_+^2)}\sum_{k=1}^{\infty}\frac{k^{2k-2}}{(2k)!}B^k  \, , \,\,\,\, 
\end{eqnarray}
with the $k$-th cumulant scaling as $L^{(k-2)}$ as can be explained by 
 the domain wall picture   for $\rho_a\ll 1$  \cite{DWT,LazarescuMallick}.
  We note 
 that this is  the only case where the statistics of the current depend
  on both the system size and on  the boundary parameters at the large $L$ limit.

  We have obtained exact formulas for the current  statistics 
  of  the open exclusion process in contact with
  two reservoirs. Our  results are valid for arbitrary sizes and values of the
  parameters and  have been tested by  precise DMRG
  computations in various regions of
  the phase diagram. They could  also  be used as benchmarks to test 
   alternative computational algorithms  \cite{Kurchan}.
 In the limit of large size systems, the
  asymptotic behavior of the large deviation function is derived  in all regions
  of the phase diagram as long as the  asymmetry $(1-q)$ remains finite.
  The diffusive limit $q \to 1$ represents  an important open  analytical problem
  and  the exact formulas  should coincide  with the predictions of  
  macroscopic fluctuation theory \cite{Bertini,BoDerr}. 
  We have  used an extension  of the 
  matrix Ansatz, that was  introduced for   multispecies
  exclusion models \cite{Multi}. The 
   relation between  multispecies models and current fluctuations
  (and also    between  open  and  periodic systems)  is mysterious as  no direct
  mapping is  known. We believe that our  results
  should be derivable from   Bethe Ansatz
  for a  spin  chain  with non-diagonal boundaries,
  but the corresponding
   Bethe equations have not yet been  derived  \cite{deGierNew}. 
  Finally, we emphasize that  the matrix representation
  given here  contains all the information
  about the density  profiles that generate atypical currents:  the precise calculation
 of these profiles represents a challenging  open question \cite{DerridaCairns}.


\begin{thebibliography}{99}

\bibitem{DerridaCairns}   B. Derrida,   J. Stat. Mech.: Theor. Exp.  P07023  (2007);
 J. Stat. Mech. P01030 (2011).

\bibitem{Gallavotti} D. J. Evans, E. D. G. Cohen and G. P. Morriss,
 Phys. Rev. Lett. {\bf 71}, 2401 (1993). G. Gallavotti and E. D. G. Cohen,  Phys. Rev. Lett.
{\bf 74}, 2694 (1995). C. Jarzynski,  Phys. Rev. Lett. {\bf 78}, 2690  (1997).

\bibitem{LebSpohn}  J. L. Lebowitz and H. Spohn,  J. Stat. Phys. {\bf 95}, 333 (1999).

\bibitem{Zia}    B.~Schmittmann and R.~K.~P. Zia, 
  in {\em Phase Transitions and Critical Phenomena vol 17.}, C. Domb and
 J.~L.~Lebowitz Ed., (San Diego, Academic Press,1995).

\bibitem{Schutzrev}    G.~M.~Sch\"utz,
 in {\em Phase Transitions and Critical Phenomena vol 19.}, C. Domb and
 J.~L.~Lebowitz Ed., (Academic Press, San Diego,2001).

\bibitem{Spohn}  H. Spohn, 1991,
{\em Large scale dynamics of interacting particles},
 (Springer-Verlag, New-York).

\bibitem{Bertini}  L. Bertini, A. De Sole, D. Gabrielli,
  G. Jona-Lasinio and C. Landim,
  Phys. Rev. Lett.  {\bf 87}, 040601 (2001);   Phys. Rev. Lett.  {\bf 94}, 030601 (2005). 

 \bibitem{Bodineau}  T.~Bodineau and  B.~Derrida, 2004, 
 Phys. Rev. Lett.  {\bf 92}, 180601;  2005,  Phys. Rev. E {\bf 72}, 066110.

\bibitem{DLeb}
 B. Derrida, J. L. Lebowitz,  Phys. Rev. Lett.  {\bf 80}, 209 (1998); 
 B. Derrida and C. Appert,  J. Stat. Phys. {\bf 94}, 1 (1999).
 B. Derrida, M. R. Evans,  J. Phys. A: Math. Gen. {\bf 32}, 4833  (1999)

\bibitem{DLSpeer}  B. Derrida, J. L. Lebowitz and E. R. Speer, 2002
 Phys. Rev. Lett.  {\bf 89}, 030601.

\bibitem{Gorissen} M. Gorissen, J. Hooyberghs and C. Vanderzande,  Phys. Rev. E
 {\bf 79} 020101 (2009);  M. Gorissen and C. Vanderzande,  J. Phys. A: Math. Theor. {\bf 44} 115005 (2011).

\bibitem{Kurchan} C. Giardina,  J. Kurchan and L. Peliti,  Phys. Rev. Lett.
  {\bf 96} 120603 (2006).  C. Giardina,  J. Kurchan,  V. Lecomte  and  J. Tailleur,
 J. Stat. Phys.  {\bf 145}, 787  (2011).
 P. I. Hurtado and P. L. Garrido,  Phys. Rev. Lett. {\bf 102}, 250601 (2009).

\bibitem{Takeuchi} K. A. Takeuchi and M. Sano, Phys. Rev. Lett. {\bf 104},
230601 (2010).

\bibitem{Touchette} H. Touchette, Phys. Rep. {\bf 478}, 1 (2009).

\bibitem{DerridaRep}
 B. Derrida, 1998, 
 Phys. Rep.  {\bf 301}, 65.

\bibitem{KLS} S.  Katz,  J. L. Lebowitz and H. Spohn,  J. Stat. Phys.
 {\bf 34},  497 (1984).

\bibitem{Varadhan}  S. R. S. Varadhan, 1996, {\em  The complex story of simple
exclusion}, in  It$\hat{\rm o}$ stochastic calculus and probability theory, 385 
(Springer, Tokyo). 

\bibitem{Chou} T. Chou and D. Lohse,   Phys. Rev. Lett.  {\bf 82}, 3552 (1999).
  T. Chou, K. Mallick and R. K. P. Zia, 
 Rep. Progr.  Phys.  {\bf 74},  116601 (2011).

\bibitem{Johansson} K. Johansson,  Comm. Math. Phys.  {\bf 209},  437  (2000)
 T. Kriecherbauer and J. Krug,  J. Phys. A: Math.  Theor. {\bf 43}, 403001 (2010).
 I. Corwin,  Random Matrices, Theory and Appl.    {\bf 1},  (2012). 


\bibitem{KPZ} M. Kardar, G. Parisi and Y.-C. Zhang,  Phys. Rev. Lett.
  {\bf 56}, 889 (1986).
  T. Halpin-Healy, Y.-C.~Zhang,  Phys. Rep.  {\bf 254}, 215 (1995).
 
\bibitem{Krug}  J. Krug,   Phys. Rev. Lett. {\bf 67}, 1882 (1991).

\bibitem{SasamSpohn} T. Sasamoto, J. Stat. Mech.: Theor. Exp.  P07007 (2007).
 T. Sasamoto and H. Spohn,  Phys. Rev. Lett. {\bf 104}, 230602 (2010).
  G. Amir, I. Corwin and J. Quastel, Comm. Pure Appl. Math.  {\bf 64},
   466 (2011).
  P. L.  Ferrari, J. Stat. Mech.: Theor. Exp.  P10016  (2011).

\bibitem{Dhar} D. Dhar,  Phase Transit. {\bf 9}, 51 (1987).
 L.-H.Gwa and H. Spohn,  Phys. Rev. Lett.  {\bf 68}, 725 (1992). 
 D. Kim, Phys. Rev. E {\bf 52} 3512 (1995).  J.~de~Gier, F.~H.~L. Essler,
 Phys. Rev. Lett.  {\bf 95}, 240601 (2005).

\bibitem{Sylvain4} S. Prolhac, J. Phys. A: Math. Theor. \textbf{43}, 105002 (2010).

\bibitem{TracyWidom} C.A.  Tracy, H.  Widom,  Comm. Math. Phys.  {\bf 279},
 815 (2008);  J. Math. Phys. {\bf 50}, 095204 (2009). 

\bibitem{SchutzDet}   G.~M.~Sch\"utz,   J. Stat. Phys. {\bf 88}, 427 (1997).
A. Rakos and  G.~M.~Sch\"utz,  J. Stat. Phys. {\bf 118}, 511 (2005).

\bibitem{DeDoMuk}  B. Derrida, E. Domany and D. Mukamel,
   J. Stat. Phys. {\bf 69},  667 (1992). 
 G.~M.~Sch\"utz and E. Domany,  J. Stat. Phys. {\bf 72}, 277 (1993).

\bibitem{DEHP}
 B. Derrida, M.~R.~Evans, V. Hakim and  V. Pasquier, 
J. Phys. A: Math. Gen. {\bf 26}, 1493 (1993).



\bibitem{MartinReview}  R.~A.~Blythe and  M. R. Evans, 
 J. Phys. A: Math. Theor. {\bf 40}, R333 (2007).


\bibitem{deGierNew}
  J.~de~Gier, F.~H.~L. Essler, Phys. Rev. Lett.  {\bf 107}, 010602 (2011). 


 \bibitem{BoDerr}  T.~Bodineau and  B.~Derrida,
 J. Stat. Phys. {\bf 123}, 277 (2006). 

\bibitem{DWT} M. Dudzinski and G. M. Sch\"utz,
J.Phys. A: Math. Gen. {\bf 33}, 8351 (2000).

\bibitem{Doucot}  B. Derrida, B. Dou\c{c}ot and P.-E. Roche, 2004,
 J. Stat. Phys. {\bf 115} 717.

\bibitem{SasaPasep1}  T. Sasamoto,  J. Phys. A: Math. Gen.
 {\bf 32}, 7109 (1999). M. Uchiyama, T. Sasamoto and M. Wadati,  
 J. Phys. A: Math. Theor. {\bf 37}  4985 (2004);  M.  Uchiyama  and M. Wadati,
J.Nonlinear Math. Phys. {\bf 12}, 676 (2005).

 \bibitem{DEMal}  B. Derrida, M. R. Evans and  K. Mallick, 
  J. Stat. Phys. {\bf 79}, 833 (1995).


\bibitem{Multi} S. Prolhac, M. R. Evans and K. Mallick,
 J. Phys. A: Math. Theor. {\bf 42} 165004 (2009).

\bibitem{LazarescuMallick} A. Lazarescu and K. Mallick,
  J. Phys. A: Math. Theor. {\bf 44} 315001 (2011).


\bibitem{LazAl} A. Lazarescu and al., {\it in preparation}.


\bibitem{White}  S. R. White,  Phys. Rev. Lett.  {\bf 69} 2863 (1992); 
 Phys. Rep. {\bf 301} 187 (1998). 
 U. Schollw\"ock, Ann. Phys. (NY) {\bf 326}, 96 (2011).

\end{thebibliography}
\end{document}